\documentclass[10pt,conference,twocolumn]{IEEEtran}
\usepackage{geometry} 
\title{Challenges to be addressed for realising an Ephemeral Cloud Federation}

\begin{document}

\author{\IEEEauthorblockN{Emanuele Carlini, Massimo Coppola, Patrizio Dazzi}
\IEEEauthorblockA{ISTI-CNR\\
Pisa, Italy \\
Email: name.surname@isti.cnr.it}
\and
\IEEEauthorblockN{Matteo Mordacchini}
\IEEEauthorblockA{IIT-CNR\\
Pisa, Italy\\
Email: name.surname@iit.cnr.it}
}

\maketitle

\begin{abstract}
This paper sketches the challenges to address to realise a support able to achieve an Ephemeral Cloud Federation, an innovative cloud computing paradigm that enables the exploitation of a dynamic, personalised and context-aware set of resources. 
The aim of the Ephemeral Federation is to answer to the need of combining private data-centres with both federation of cloud providers and the resource on the edge of the network. 
The goal of the Ephemeral Federation is to deliver a context-aware and personalised federations of computational, data and network resources, able to manage their heterogeneity in a highly distributed deployment, which can dynamically bring data and computation close to the final user.
\end{abstract}


\section{Introduction}

Cloud Computing has transformed the IT by providing computing and storage as on-demand utility services according to the pay per use model.
Large IT behemoths provided their spare resources for renting to other private enterprises and individuals.
With the evolution of services and applications that can be brought on the cloud, a single cloud solution cannot provide the heterogeneity and functionality required for many business solution. Therefore, the initial concept of cloud computing has rapidly evolved into multi-cloud environment, which gather together multiple and heterogeneous cloud data-centres and service providers. 

The earlier protagonist of today's multi-cloud era are the so called \emph{Cloud Federations}. Cloud federation brought the cloud computing to the next level, realising much more than inter-cloud interoperability~\cite{ISGC2012,smartfed2013}, rather providing an unified view on an heterogeneous pool of resources while using a single access point to control applications. In the Cloud Federation model a number of cloud providers voluntary join their resources to collaboratively increase their market and to achieve scale economy that would have been outside their reach. 

Due to the innovations in virtualisation solution and approaches, the original federation concept has evolved over the time in more complex and functionality-rich paradigms. One of these paradigms is the \emph{hybrid-cloud}, in which private resources of an organisation are combined with public resources belonging to (federations of) cloud data-centres. 
Another relevant paradigm is the one of \emph{mobile-cloud}, in which mobile devices at the edge of the network share the same execution environment of large cloud data-centre, in principle allowing a seamless migration of computation back and forth on both environments.

These two paradigms offer two views of the tendency of moving the computation toward cloud resources. On one side there is the possibility to burst toward public resources when internal resource are not enough or to cope with peaks of utilization. On the other side, the devices at the edge of the network participate to a collective realisation of services, and they offload computation and data in the cloud to support these demanding applications that would have been outside their computational capacities otherwise.
Unfortunately, the evolution of hybrid- and mobile- cloud followed two different paths, and many approaches developed for one cannot be used as they are for the other. 
Therefore, there is the need of a unifying approach that would combine hybrid and mobile cloud in a common execution environment. This vision would open up new possibilities for cloud based applications in terms of cost-effectiveness, scalability and ability to use many kinds of heterogeneous resources. 

In order to realise this vision we introduce the concept of Ephemeral Cloud Federation, a cloud environment that realises more than the convergence of hybrid and mobile cloud paradigms, but offer the customers a dynamic, personalised and context-aware view over a pool of resources. The remaining of this paper is organised as follows: in Section~\ref{sec:background} is presented the current background related to approaches involving the management of multiple clouds. Section~\ref{sec:challenges} defines the current open challenges in the field. Then, Section~\ref{sec:ECF} introduce the concepts underpinning the Ephemeral Cloud Federation. Finally, in Section~\ref{sec:conclusion} are drawn our conclusions.

\section{Background}\label{sec:background}

Many approaches for Cloud Federation aimed to the creation of unified access points that act as gateway toward the whole federation resources. 
Approaches such as Contrail~\cite{carlini2011cloud} and Intercloud~\cite{buyya2010intercloud} employ the concept a centralised and structured entity that works as the controller of the federation. In particular, Contrail focuses both on the vertical and horizontal integration of multiple cloud providers, organising the enforcement of QoS by the definition of federation-level SLAs, which also driver the resource selection process and can be mapped on single cloud providers SLAs. Conversely, Intercloud's idea of Cloud Federation is realised via the definition of a common marketplace in which applications are negotiated among brokers and cloud providers.
Other approaches, such as RESERVOIR~\cite{rochwerger2009reservoir} and OPTIMIS~\cite{ferrer2012optimis}, provided a more distributed view in the landscape of Cloud Federation.
%

More recently, cloud federations architectures and approaches have been extended to support enterprises that have sensible data and applications that must be kept in local premises, giving birth to the \emph{hybrid} cloud model. 
In this model, private resources (e.g. a local cloud datacenter) of an enterprise are mingled with public cloud resources (including Cloud Federation), with the aim to combine the cost-effectiveness of public cloud with the control of private cloud~\cite{kashef2011cost}. Hybrid cloud is typically used in the context of Cloud bursting, that is the exploitation of public cloud to support peak workload. Many of the current cloud management framework, such as OpenNebula~\cite{milojivcic2011opennebula} and OpenStack~\cite{sefraoui2012openstack}, offer support to cloud bursting from private to public cloud. The utilisation of cloud bursting in hybrid cloud has been exploited in several domains, under to form of services offered to enterprises, in which the knowledge of the application and the datacenters is complete and static. However, these systems can take up to hours to ``burst'', and the decision to when is convenient to burst is left to the service owner~\cite{guo2014cost,6973755}. To the best of our knowledge, only few works tackled the problem of cloud bursting in hybrid cloud aiming scalability and generality~\cite{mateescu2011hybrid,sotomayor2009virtual}.

%
%
A popular mobile cloud approach is the one based on \emph{cloudlet}, originally proposed by Satyanarayanan \textit{et al.}~\cite{satyanarayanan2009case}. In this model, mobile devices offload their computation to \emph{cloudlets}, which are relatively small computational units connected with the full blown remote cloud server. Cloudlets are deployed locally to the devices and often placed in common and crowded areas to achieve physical proximity with mobile devices. This aspect provides devices with low latency and high bandwidth connections, thereby allowing an interactive response for highly demanding applications. 

\section{Challenges}\label{sec:challenges}
Current trends in cloud computing show a prominent interest in hybrid cloud solutions, where users, and their respective applications, can take advantage from a joint usage of public clouds  and users' local/private resources. This kind of solution is proving to be very effective in enabling companies to keep the most critical portions of data and applications on their own datacentres and, at the same time, being able to move the less (or not) critical applications and data to the cloud. However, this is actually done at the price of a very carefully designed data decomposition and application orchestration. 

In addition to this, another growing trend in cloud computing regards the evolution towards more decentralised, edge-based approaches. Users are increasingly accessing cloud services from mobile devices in scenarios where other devices static and mobile available in specific physical areas, generate raw data and provide distributed cloud computing functions. In order to fully exploit these resources, traditional cloud solutions based on remote public clouds may not necessarily be the best response in such a dynamic and local context. In contrast, locating the computing functions towards the edge of the network has the advantage of make it easier to support services with low latency requirements and to facilitate real-time interactions between the nodes in the architecture. 
Services based on local data may be they themselves of local nature such as, for example, spontaneous interactions between co-located users or proximity-based services. Therefore, local data do not necessarily need to be transferred to a global cloud platform to be processed, as might be relevant only to the users close to where they are generated. Aggregated information, on the other hand, might be even more useful for users distant from where local data is generated. Last but not least, it might even not be possible to move data available on users personal devices, due to privacy constraints. However, supporting cloud services and applications in this kind of context is extremely challenging from the perspective of services and resources orchestration due to the high dynamicity and the intrinsically  distributed and localised nature of the environment. 

A next generation cyber-infrastructure capable of targeting the above-mentioned needs, arising from cloud users and novel cloud applications, needs to provide an effective and efficient deployment on a widely distributed and heterogeneous cloud federation infrastructure, that should encompass both hybrid cloud and edge resources located in the network. 
In the following, we highlight the main challenges related to the deployment of a novel cloud infrastructure that federates public, private and edge resources.

In our envisioned Ephemeral cloud federation, applications and data can be decomposed. Therefore, different pieces of data and parts of the overall application can be potentially placed in different types of cloud resources. To overcome current hybrid cloud solutions limitations, the decomposition and the deployment of the application should be made in a seamless way, rather then in an ad-hoc manner. Therefore, the main challenge is to allow the system to optimally decide where to deploy the different parts of an application and its required data, limiting the direct intervention of the user to the definition of policies and requirements that should guide the deployment. 

This challenge is made more difficult by the presence of edge resource. In fact, the degree and the way in which the data and computation is divided and moved from/to the edge devices to/from the other cloud resources should be adaptively decided based on the context and the kind of requests of the user. Thus, the main challenge is to dynamically determine which are the most suitable resources that are needed to fulfill the user request. These  resources are personalised to each single user and the application she is requesting. In fact, each different user has access to her own private resources, that are part of a hybrid cloud. Every distinct user has also different security and privacy needs, that depends on the type of application that is requested. This fact impacts the possible choice between public and private resources, and affect the set of edge resources that could be recruited for running the service. As a further personalization, this set should be necessarily derived from the actual context surrounding the user. Moreover, this set is also application-dependant, since the requirements of the application limits the possible type of edge resources that could be considered. The application deployment should then be dynamically adapted depending on the actual user context (data, resources, location, etc.), while matching the need for robustness, trustworthiness and performance. 

Dynamically detecting the context of the user and the resources at the edge require to face other challenges. In fact, another fundamental challenge is to determine how data can be reliably and efficiently collected and disseminated at the edge, in order to allow to effectively deploy the application. Another fundamental challenge is how to support simple services, such as resource discovery or continuous data provisioning at the edge. This is a heterogeneous environment where devices may have extremely limited resources in terms of energy, storage and computing. In this case, the design of the system should consider f efficient techniques that allow to face the challenge to perform service discovery in a distributed manner while ensuring low latency and fast response to network dynamics (link/node failures, mobility), as well as low energy consumption.

Once the application and its required data are distributively deployed, the system should support support the execution of the service in a such a context, facing the challenge of how to efficiently orchestrate distributed the various service components in a system that federates public clouds as well as the user's personal private cloud and edge devices. Problems such as how to compose multiple resources, how to select the sequence of service components to execute, how to optimally migrate components and the data needed to execute a service inside this scenario should be considered.

\section{Ephemeral Cloud Federation}\label{sec:ECF}
As the aforementioned challenges outline, there is a call for paradigm switch to evolve the actual cloud technology spectrum. As matter of facts, to overcome the limitations affecting the actual solutions there is a need for models able to encompass the edge and hybrid-cloud computing approaches. The necessity of creating a dynamic, context-aware and personalised federations of computational, data and network resources, able to manage their heterogeneity in a highly distributed deployment, consisting in public and private data-centres as well as in a very different types of resources localised at the edge of the network, which can be dynamically involved into the computation to bring data and computation close to the ``consumer''.
To address these needs we propose an innovative concept that we identify with the term ``Ephemeral Cloud Federation''. It consists in an effective and efficient enabling approach allowing the deployment of cloud applications on a widely distributed and heterogeneous cloud federation infrastructure, involving both traditional cloud- and edge-resources located at the network. The approach will provide mechanisms to dynamically adapt application deployment depending on the actual context (data, resources, location, etc.) and match the need for robustness, trustworthiness and performance. Starting from initial requirements on applications and data, and by exploiting monitoring and context information, the Ephemeral Cloud Federation is able to define, enact and dynamically optimise application deployment and runtime management plans. 
The approach will provide complete control over deployment choices (e.g. decide where the data should be stored and processed, and how to secure them). It will then enable the application deployment without the burden of managing all the necessary low-level activities. The Ephemeral Cloud Federation will dynamically recruit and organize heterogeneous resources from multiple sources, blending various kinds of devices and in-house resources with public clouds, mobile devices, enterprise data-centres and special hardware such as GPU or SSD. 
It will provide an integrated API and the corresponding solutions to support brokering, runtime discovery and provisioning of the complex set of distributed devices composing it: a wide set of different types of resources ranging from public clouds to specialised resources standing at the extreme edge of the network. 
To this end, the Ephemeral Cloud Federation requires the development and the operation of peculiar technologies to ease the extension of the existing platform to support new protocols and resources, that will also contribute to standardisation efforts in the area of provider interoperability. It will also introduce mechanisms to operatively realise the federation, addressing the issues of distributed access control management, privacy, security and performance management that affect both the federation decisions (i.e. with respect to specific environments and administrative domains rules) and the service offerings (i.e. with respect to location-specific characteristics). 
%

\subsection{Dynamic, Personalised and Context-aware resources}

The Ephemeral Cloud Federation enables the exploitation of a dynamic, personalised and context-aware set of resources. The Ephemeral Cloud Federation is composed by a set of private cloud data-centres and edge devices that a given user, when using a certain application, can access and exploit, depending on the actual context. The Ephemeral Cloud Federation is said to be ephemeral as there is no a single, unique, consolidate view on the various resources of the federation, but such view is personalised according to the user and depends on multiple factors. 
%
These factors include the user preferences, the application requirements, the place where the application is executed, the actual situation and the moment in time when the application is executed, and the dynamic availability of resources during the application execution. 


The Ephemeral Cloud Federation need to provide high performances, high dynamicity and high scalability, which can be achieved by means of a tight interaction between the orchestration of the application and the organisation and management of the resources executing the application. To this respect, the ephemeral cloud federation can represent a good fit as the personalisation inherently defines a strict relationship among application and federated resources. 
%
Applications require to be secured, as their business logic and data (or a part of it) can be private, and/or subject to no disclosure and non-authorised accesses. In this sense, the ephemeral cloud federation aims at exploiting current, and design new, technologies and methodologies to enforce trustworthiness, robustness and performance for data-intensive applications within the ephemeral cloud federation. This is achieved by means of an advanced support to application and data orchestration, allowing users to define policy about data placement and distribution eventually and automatically realised by the Ephemeral Cloud Federation platform.

All these aspects concur to make the set of resources that an application can exploit very dynamic and flexible. This makes the ephemeral cloud federation a very powerful and flexible computation platform for a wide range of applications. 

\subsection{Services on the Ephemeral Cloud Federation}
Services are submitted to the Ephemeral Cloud Federation along with a representation of their computational and security requirements. The Ephemeral Cloud Federation provides an advanced support for application orchestration, allowing customers to easily drive the placement of applications and data according to their preferences and needs and matching the requirements of a robust, secure and performant infrastructure. 
Application-related information and requirements are complemented with the context characterising the actual user. This allows the Ephemeral Cloud Federation to define an effective deployment plan, which is enforced and continuously observed by means of an advanced monitoring subsystem, to be dynamically fine-tuned by the runtime application management support. 

From an operative viewpoint, the allocation process is conducted by the Ephemeral Cloud Federation at different stages, at different levels of abstraction, and at different places. Applications, their actual users and the associated contexts are analysed to derive an initial, high-level allocation plan. Such plan can be locally performed by any access point to the federation. It takes into consideration a coarse representation of all the resources available in the federation. Basically, it consists in defining where the different instances of the services composing the application should be placed, in terms of data-centres, or overall set of resources (e.g., a certain public cloud, a certain set of edge resources). Then, it takes place a finer scheduling/mapping process targeting the specific resources selected for hosting the application, e.g. intra-cluster mapping. This completes the initial application allocation process. After, it is started the monitoring process. The purpose of this activity is to keep informed the application runtime management, devoted to the continuous optimisation and management of the applications. Such process takes into account the information deriving from the observations and monitoring of the applications at run-time, in order to provide the necessary management plans resulting in specific application re-configuration operations, such as migration as well as horizontal- and vertical-elasticity.

\section{Conclusion}\label{sec:conclusion}

This position paper details the idea and the concepts behind the Ephemeral Cloud Federation, an innovative multi-cloud paradigm that enables the exploitation of a dynamic, personalised and context-aware set of resources.  We believe that this paradigm represents the conjunction between two similar but still separated worlds, namely Hybrid-clouds and Mobile clouds. 
In this context, we have outlined the main challenges that arise to execute applications on a platform made of the combination of cloud datacenters and the devices on the edges. In order to overcome these challenges, we have designed a preliminary version of the architecture of the ephemeral federation, and described the main components and their role in the ephemeral cloud federations. 

\nocite{dazzi2005java,Dazzi09LSDSIRp2pclustering}

\end{document}